\documentstyle[aps,prl,epsfig,twocolumn]{revtex}

\def\be{\begin{equation}}
\def\ee{\end{equation}}
\def\bea{\begin{eqnarray}}
\def\eea{\end{eqnarray}}

\begin{document}
\draft

\title{Entanglement of atoms via cold controlled collisions}

\author{
D. Jaksch,$^{1}$ H.-J.  Briegel,$^{1}$ J.I. Cirac,$^{1}$ C. W. Gardiner,$^{2}$ and P. Zoller$^{1}$}

\address{$^{1}$Institut f\"ur Theoretische Physik, Universit\"at Innsbruck,
Technikerstrasse 25, A--6020 Innsbruck, Austria.\\
$^{2}$School of Chemical and Physical Sciences, Victoria University, Wellington, New Zealand.
}

\date{\today}

\maketitle

\begin{abstract}
We show that by using {\em cold controlled collisions} between two 
atoms one can achieve conditional dynamics in moving trap potentials. 
We discuss implementing two qubit quantum--gates and efficient creation
of highly entangled states of many atoms in optical lattices.
\end{abstract}



\narrowtext


The controlled manipulation of entangled states of $N$-particle systems is fundamental to the study of basic aspects of quantum theory \cite{Bell,Gr89}, and provides the basis of applications such as quantum computing and quantum communications \cite{book,Gatepapers}. Engineering entanglement in real physical systems requires precise control of the
Hamiltonian operations and a high degree of coherence. 
Achieving these conditions is extremely demanding, and only a
few systems, including trapped ions, cavity QED and NMR, have been identified as possible candidates to implement quantum logic in the laboratory \cite{book}. On the other hand, in atomic physics with {\em neutral atoms} recent advances in cooling and trapping have led to an exciting new generation of experiments with Bose condensates \cite{BECreview}, experiments with optical lattices \cite{Hansch}, and atom optics and interferometry. 
The question therefore arises, to what extent these new experimental possibilities and the underlying physics can be adapted to provide a new perspective in the field of experimental quantum computing. 

In this Letter we propose coherent cold collisions as the basic mechanism
to entangle neutral atoms. The picture of {\em atomic collisions} as {\em coherent interactions}  has emerged during the last few years in the studies of  Bose Einstein condensation (BEC) of {ultracold  gases}. In a field theoretic language these interactions correspond to  Hamiltonians which are quartic in the atomic field operators, analogous to Kerr nonlinearities between photons in quantum optics. 
By storing ultracold atoms in arrays of microscopic potentials provided, for example, by optical lattices these collisional interactions can be controlled via laser parameters. 
Furthermore, these nonlinear atom-atom interactions can be large \cite{Mott}, even for interactions between individual pairs of atoms, thus providing the necessary ingredients to implement quantum logic.

Let us consider a situation where two atoms $|a\rangle$ and $|b\rangle$
are trapped in the ground states $\psi^{a,b}_0$ of two potential wells
$V^{a,b}$. Initially, at time $t=-\tau$, these wells are centered at
positions $\bar x^a$ and $\bar x^b$, sufficiently far apart 
(distance $d=\bar x_b -\bar x_a$) so that the particles do not interact.
The positions of the potentials are moved along trajectories $\bar
x^a(t)$ and $\bar x^b(t)$ so that the wavepackets of the atoms overlap
for certain time, until finally they are restored to the initial position
at the final time $t=\tau$. This situation is described by the
Hamiltonian
\be
\label{Hamil}
H \!= \! \sum_{\beta=a,b} \left[ \frac{(p^\beta)^2}{2m} + V^\beta\left(x^{\beta}\!-\!
\bar x^\beta(t)\right) \right] + u^{\rm ab}(x^a\!-\!x^b).
\ee
Here, $x^{a,b}$ and $p^{a,b}$ are position and momentum operators,
$V^{a,b}\left(x^{a,b}-\bar x^{a,b}(t)\right)$ describe the displaced trap potentials and 
$u^{ \rm ab}$ is the atom--atom interaction term. 

Ideally, we would like to implement the transformation from $t=-\tau$ to $t=\tau$
\be
\label{transf}
\psi^a_0(x^a \!-\!\bar x^a) \psi^b_0(x^b\!-\!\bar x^b) \rightarrow
e^{i\phi} \psi^a_0(x^a\!-\!\bar x^a) \psi^b_0(x^b\!-\!\bar x^b),
\ee
where each atom remains in the ground state of its trapping potential and
and preserves its internal state. The phase $\phi$ will contain a
contribution from the interaction (collision). Transformation
(\ref{transf}) can be realized in the {\em adiabatic limit}, \cite{GalPas} whereby we
move the potentials so that the atoms remain in the
ground state. In the absence of interactions ($u^{\rm ab}=0$)
adiabaticity requires $|\dot{\bar x}^{a,b}(t)| \ll
v_{\rm osc} \quad \forall t$, where $v_{\rm osc} \approx
a_0\omega$ is the rms velocity of the atoms in the vibrational ground
state, $a_0$ is the size of the ground state of the trap potential, and
$\omega$ is the excitation frequency. The phase $\phi$ can be easily
calculated in the limit $|\ddot{\bar x}^{a,b}(t)| \ll v_{\rm osc}/\tau$.
In this case, $\phi=\phi^a + \phi^b$ where
\be
\phi^{a,b} = \frac{m}{2 \hbar} \int_{-\tau}^\tau dt \dot{\bar x}^{a,b}(t)^2,
\ee
are the {\em kinetic phases}. In the presence of
interactions ($u^{\rm ab} \ne 0$), we define the time--dependent energy shift due to the interaction
as
\be
\label{deltaE}
\Delta E(t)= \frac{4\pi a_s\hbar^2}{m}\int dx \prod_{\beta=a,b}|\psi^\beta_0
\left(x-\bar{x}^\beta(t)\right)|^2,
\ee
where $a_s$ is the $s$--wave scattering length. We assume that: (i)
$|\Delta E(t)|\ll \hbar \omega$ so that no sloshing motion is excited; 
(ii) $|\dot{\bar x}^{a,b}(t)| \ll v_{\rm osc}$
(adiabatic condition); (iii) $v_{\rm osc}$ is sufficiently small for the
zero energy $s$--wave scattering approximation to be valid \cite{note}. In
that case, (\ref{transf}) still holds with
$\phi=\phi^a + \phi^b + \phi^{\rm ab}$, where 
\be
\label{phicol}
\phi^{\rm ab} = \frac {1}{\hbar} \int_{-\tau}^\tau dt \Delta E(t),
\ee
is the {\em collisional phase}.

In the case of quasi--harmonic potentials (as is realized with optical
potentials, see below) (\ref{transf}) still holds, even
in the non--adiabatic regime, i.e.~at higher velocities.
For harmonic traps one can solve exactly the evolution for $u^{\rm ab}=0$, 
and  identify the condition for adiabaticity:
\be
\label{adia}
\left| \int_{-\tau}^t \dot{\bar x}^{a,b}(t') e^{i\omega t'} dt' \right| \ll a_0, 
\quad \forall -\!\tau\ge t\ge \tau.
\ee
This is consistent with condition (ii). Actually, for (\ref{transf}) to hold the
inequality (\ref{adia}) only has to be fulfilled for $t=\tau$ (and not
for all times). This means that the particle need not be in
the ground state of the moving potential at all times, but only at the
final time. The phase $\phi=\phi^a + \phi^b$, as well as the wave
functions $\psi^{a,b}(x^{a,b},t)$, can be calculated exactly  \cite{GalPas}.
For $u^{\rm ab}\ne 0$ if conditions (i) and (iii)
are satisfied, then (\ref{transf}) is still valid. There
is an additional phaseshift (with respect to $u^{\rm ab}=0$) $\phi^{\rm
ab}$ which is given by Eqs.\ (\ref{deltaE},\ref{phicol}) with the
replacement $\psi^{a,b}_0(x^{a,b}-\bar x^{a,b}(t)) \rightarrow
\psi^{a,b}(x^{a,b},t)$. It is also straightforward to generalize these
results to the case in which the trap frequency changes with time \cite{Shlyap}.

So far, we have shown that one can use cold collisions as a
coherent mechanism to induce phase shifts in two--atom interactions in a
controlled way. Our goal is now to use these interactions to 
implement conditional dynamics. We consider two atoms $1$ and
$2$, each of them with two internal levels $|a\rangle_{1,2}$ and
$|b\rangle_{1,2}$. We will use the superscripts $\beta={a,b}$ and the subscripts
$j={1,2}$ to label the internal levels and atoms, respectively. Atoms in the internal 
state $|\beta\rangle_j$ experience a potential $V^\beta_j$ which is
initially ($t=-\tau$) centered at position $\bar x_j$. We assume that we can
move the centers of the potentials as follows (Fig.\ 1): $\bar
x^\beta_j(t)=\bar x_j+ \delta x^\beta(t)$. The trajectories $\delta x^\beta(t)$ are 
chosen in such a way that $\delta x^\beta(-\tau)=\delta x^\beta(\tau)=0$ and the 
first atom collides with the second one only if they are in states $|a\rangle$ and
$|b\rangle$, respectively ($|\bar x_1^b(t)-\bar x_2^a(t)|\gg a_0$ $\forall t$). This choice is 
motivated by the physical implementation considered below. The fact that $\bar x_j$ 
does not depend on the internal atomic state allows one to easily change this 
internal state at times $t=\pm \tau$ by applying laser pulses. If the conditions
stated above are fulfilled, depending on the initial internal atomic states we 
have the following table for this process:
\bea
\label{gate}
|a\rangle_1 |a\rangle_2 &\rightarrow& 
  e^{i2\phi^a} |a\rangle_1 |a\rangle_2,\nonumber\\
|a\rangle_1 |b\rangle_2 &\rightarrow& 
  e^{i(\phi^a+\phi^b+\phi^{\rm ab})} |a\rangle_1 |b\rangle_2,\nonumber\\
|b\rangle_1 |a\rangle_2 &\rightarrow& 
  e^{i(\phi^a+\phi^b)} |b\rangle_1 |a\rangle_2,\nonumber\\
|b\rangle_1 |b\rangle_2 &\rightarrow& 
  e^{i2\phi^b} |b\rangle_1 |b\rangle_2,
\eea
where the motional states remain unchanged. 
The kinetic phases $\phi^\beta$ and the collisional phase
$\phi^{\rm ab}$ can be calculated as stated above. 
We emphasize that the $\phi^\beta$ are (trivial)
one particle phases that, if known, can always be incorporated in
the definition of the states $|a\rangle$ and $|b\rangle$.

\begin{figure}[tbp]
\epsfig{file=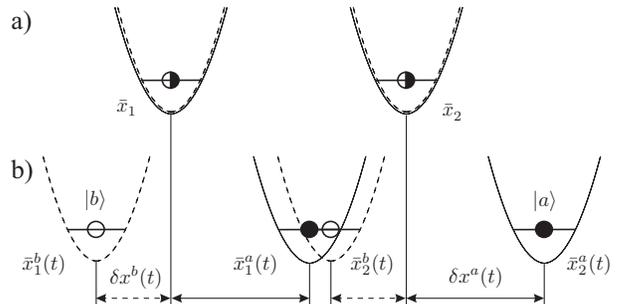,width=8cm}
\caption{Configurations at times $\pm \tau$ (a) and at $t$ (b). The 
solid (dashed) curves show the potentials for particles in the internal state 
$|a\rangle$ ($|b\rangle$), respectively. Center positions $\bar
x^\beta_j(t)$ and displacements $\delta x^\beta(t)$ as defined in the
text.}
\end{figure}

In the language of Quantum Information, transformation (\ref{gate}) corresponds to a 
fundamental two--qubit gate \cite{Gatepapers}. 
In order to illustrate to what extent the mechanism presented above is
able to perform this ideal gate, we have carried out a
numerical study in 3 dimensions. We have integrated the time--dependent Schr\"odinger
equation with the Hamiltonian (\ref{Hamil}). We have taken harmonic
potentials with various time dependent displacements $\delta x^\beta(t)$
and frequencies. Their form as well as the parameter range are motivated
by the specific implementations outlined below. The figure of merit that
we have used is the minimum fidelity, which is defined as
\be
F = \min_{\psi} \langle \tilde\psi| {\rm tr}_{\rm ext}\left(
{\cal U} |\psi\rangle\langle\psi|\otimes \rho_{\rm ext} {\cal U}^\dagger \right)
|\tilde\psi\rangle .
\ee
Here $|\psi\rangle$ is an arbitrary internal state of both atoms,
$|\tilde \psi\rangle$ is the state resulting from $|\psi\rangle$ using
the mapping (\ref{gate}). The trace is taken over motional states,
${\cal U}$ is the evolution operator for the internal states
coupled to the external motion (including the collision), and $\rho_{\rm
ext}$ the density operator corresponding to both atoms in the motional
ground state at $t=-\tau$. In the ideal case
the fidelity will be one. We have taken the potential $u^{\rm ab}$ as
proportional to a delta function and used a truncated moving harmonic
oscillator basis (10 states for each degree of freedom). The first
illustration corresponds to a fixed trap frequency $\omega^a=\omega^b$
and displacements $\delta x^b(t)=0$ and $\delta x^a(t)$ as specified in
Fig.\ 2a. In Fig.\ 2b we present a contour plot of $F$ as a function of
the parameters characterizing the displacement $\delta x^a(t)$.
The fidelity is very close to one for a
surprisingly wide range of parameters, even well into the non--adiabatic
regime. We have also studied numerically the effect of time--varying
trap frequencies (see Fig.\ 3a) and finite temperatures, so that
$\rho_{\rm ext}$ describes a thermal distribution of temperature $T$.
For temperatures $kT\alt 0.2\hbar\omega$ (where $\omega$ is the initial
trap frequency) the fidelity remains very close to one (of the order of
$0.997$), and this result is still robust with respect to changes in the
parameters. We have also included the possibility of loss due to
collisions by adding an imaginary part to the scattering length. In that
case, the fidelity is reduced (see Fig.\ 3b). 

\begin{figure}[tbp]
\epsfig{file=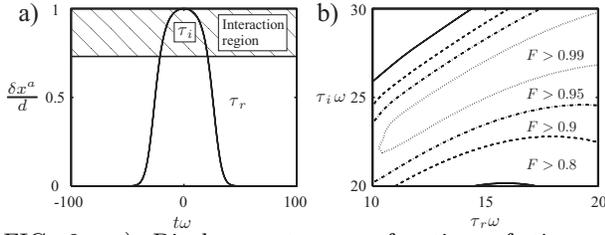,width=8cm}
\caption{a) Displacement as a function of time $\omega t$, $\delta x^a(t)/d=
\left(1+\exp(-(\tau_i/\tau_r)^2)\right)/ \left(1+\exp((t^2-\tau_i^2)/
\tau_r^2)\right)$ and $\delta x^b(t)= 0$ (see text), with 
$\tau_r=30/\omega$ and $\tau_i=20/\omega$.
The shaded region indicates where the particles interact.
b) Fidelity $F$ against rise time $\tau_r$ and interaction time $\tau_i$ for $^{87}$Rb 
with $a_s=5.1 {\rm nm}$, $\omega=2 \pi \times 100 {\rm kHz}$ and $d=10 \, a_0$.}
\end{figure}

So far we have assumed that there is one atom per potential well. In
practice, the particle number might not be controlled. Nevertheless, one
can easily generalize the above results to this case by using a second
quantized picture. For example, in the adiabatic regime we denote by
$a_i$ and $b_i$ the annihilation operators for a particle in the ground
state of the potential centered at the position $i$, and corresponding
to the internal levels $|a\rangle$ and $|b\rangle$, respectively. The
effective Hamiltonian in this regime is
\bea
H &=& \sum_{i} \left[\omega^a(t) a_i^\dagger a_i + \omega^b(t) b_i^\dagger b_i
+ u^{\rm aa}(t) a_i^\dagger a_i^\dagger a_i a_i + \right. \nonumber\\
&&\left. u^{\rm bb}(t) b_i^\dagger b_i^\dagger b_i b_i \right]
+ \sum_{i,j} u^{\rm ab}_{ij}(t) a_i^\dagger a_i b_j^\dagger b_j,
\eea
where the $\omega$'s and $u$'s depend on the specific
way the potentials are moved. This Hamiltonian corresponds to a Quantum--Non--Demolition
situation \cite{QND}, whereby the particle number can be measured non--destructively.

\begin{figure}[tbp]
\epsfig{file=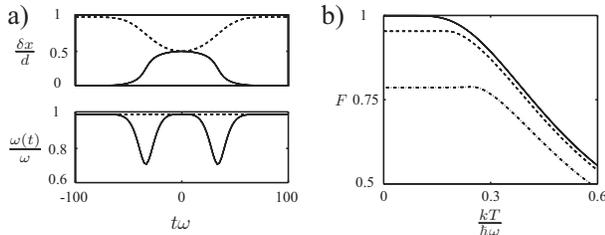,width=8cm}
\caption{a) Upper plot: Displacements $\delta x^a(t)/d$ (solid line) and $1+\delta x^b(t)/d$ 
(dashed line). Lower plot: Trap frequencies $\omega^{a}(t)/\omega$ 
(solid line) and $\omega^{b}(t)/\omega$ (dashed line). b) Fidelity $F$ against 
temperature $kT/\hbar \omega$ for $^{87}$Rb with $a_s=5.1 {\rm nm}$ (solid line), 
$a_s=(1-0.01i)\times\,5.1 {\rm nm}$ (dashed line) and 
$a_s=(1-0.05i)\times \, 5.1 {\rm nm}$ (dash-dotted line). Here
$\omega=2 \pi \times 100 {\rm kHz}$ and $d=390 {\rm nm}$.}
\end{figure}

A physical implementation of this scenario requires an
interaction which produces internal state dependent conservative trap
potentials and the possibility of moving these potentials
independently. Furthermore, the choice of the internal atomic states
$|a\rangle$ and $|b\rangle$ has to be such that they are collisionally
stable (i.e. the internal states do not change after the collision).
These requirements can be satisfied in an optical lattice. 
We consider a specific but
relevant example of alkali atoms with a nuclear spin equal to $3/2$
($^{87}$Rb, $^{23}$Na) trapped by standing waves in 3
dimensions. The internal states of interest are hyperfine levels
corresponding to the ground state $S_{1/2}$. Along the $z$ axis, the
standing waves are in the lin$\angle$lin configuration (two linearly
polarized counter-propagating traveling waves with the electric fields
forming an angle $2\theta$ \cite{Finkelstein}). The electric field is a
superposition of right and left circular polarized standing waves
($\sigma^\pm$) which can be shifted with respect to each other by
changing $\theta$,
\be
\vec E^+(z,t) = E_0 e^{-i\nu t} \left[ \vec \epsilon_+ \sin(kz\!+\!\theta) + 
\vec \epsilon_- \sin(kz\!-\!\theta)\right],
\ee
where $\vec \epsilon_\pm$ denote unit right and left circular polarization vectors,
$k=\nu/c$ is the laser wavevector and $E_0$ the amplitude. The lasers
are tuned between the $P_{1/2}$ and $P_{3/2}$ levels so that the dynamical 
polarizabilities of the two fine structure $S_{1/2}$ states corresponding to
$m_s=\pm 1/2$ due to the laser polarization $\sigma^{\mp}$ vanish, whereas
the ones due to $\sigma^\pm$ are identical ($\equiv \alpha$).
The optical potentials for these two states are
$V_{m_s=\pm 1/2}(z,\theta) =  \alpha |E_0|^2 \sin^2\left(kz\pm\theta\right)$.
We choose for the states $|a\rangle$ and $|b\rangle$ the hyperfine structure states
$|a\rangle\equiv|F=1,m_f=1\rangle$ and $|b\rangle\equiv|F=2,m_f=2\rangle$.
Due to angular momentum conservation, these states are stable under collisions (for
the dominant central electronic interaction \cite{Julienne}). The potentials
``seen'' by the atoms in these internal states are
\begin{mathletters}
\label{Vab}
\bea
V^a(z,\theta) &=& \left[ V_{m_s=1/2}(z,\theta) + 3V_{m_s=-1/2}(z,\theta)\right]/4  \\
V^b(z,\theta) &=& V_{m_s=1/2}(z,\theta).
\eea
\end{mathletters}
If one stores atoms in these potentials and they are deep enough,
there is no tunneling to neighboring wells and we can approximate
them by harmonic potentials. By varying the angle $\theta$ from $\pi/2$ to $0$,
the potentials $V^b$ and $V^a$ move in opposite directions until they completely
overlap. Then, going back to $\theta=\pi/2$ the potentials return to their original
positions. The shape of the potential $V^a$ changes as it moves.
By choosing $\theta(t)=\pi \left(1-\left(1+\exp(-(\tau_i/\tau_r)^2)\right)/
\left(1+\exp((t^2-\tau_i^2)/\tau_r^2)\right)\right)/2$ with $\tau_r=30/\omega$ and 
$\tau_i=20/\omega$, the frequencies and displacements of the harmonic
potentials approximating (\ref{Vab}) are exactly those plotted in Fig.\ 3a.
Therefore, that figure shows that under this realistic situation one can obtain 
very high fidelities. 

The scheme presented here can be used in several interesting
experiments: 
(a) One can use this method to measure the phase
shift $\phi^{\rm ab}$ and therefore determine the scattering length
corresponding to $a$--$b$ collisions. For that, one can use ideas borrowed
from Ramsey spectroscopy, namely: (i) prepare {\em all} atoms in the
superposition $(|a\rangle+|b\rangle)/\sqrt{2}$ by applying a $\pi/2$
laser pulse; (ii) shift their potentials as described above; (iii) apply
another $\pi/2$ laser pulse; (iv) detect the population of the internal
states by fluorescence. It can easily be shown that such populations
depend in a simple way on the phase shift $\phi^{\rm ab}$. One can
even determine the sign of the scattering length by applying laser pulses with pulse area different from $\pi/2$ \cite{unpub}. 
(b) In a similar way, one can also measure the
spatial correlation function by applying the second laser pulse (iii)
and population detection (iv) without moving the potential back to the
origin. This would be a
way of discriminating between Mott and superfluid phases for particles
in an optical lattice \cite{Mott}. 
(c) Apart from that, if one is able to
address individual wells with a laser, one can also perform certain
experiments which are interesting both from the quantum information and
fundamental point of view. For example, one could create an entangled
EPR pair of two particles $(|a\rangle_1|b\rangle_2 -
|b\rangle_1|a\rangle_2)/\sqrt{2}$ \cite{Bell}. This could be done by having two
atoms in neighboring wells and: (i) prepare each of them in the
superposition $(|a\rangle+|b\rangle)/\sqrt{2}$; (ii) shift their
potentials back and forth so that the phase shift $\phi^{\rm ab}=\pi$; (iii)
apply a $\pi/2$ laser pulse to the second atom. In this case one could 
test Bell inequalities and perform other
fundamental experiments. 
(d) With more than two particles one could create
higher entangled states with simple lattice operations. For example, if one has $N$ particles in $N$ 
potential wells, one could create GHZ states of the form
$(|a\rangle_1|a\rangle_2\ldots |a\rangle_N -
|b\rangle_1|b\rangle_2\ldots|b\rangle_N)/\sqrt{2}$ \cite{Gr89}. This could be easily
done, for example, if one could use a different internal state
$|c\rangle_1$ in the first atom instead of $|b\rangle_1$ \cite{foot3}.
The procedure would be as follows: (i) prepare the first atom in the
state $(|a\rangle_1 + |c\rangle_1)/\sqrt{2}$, and the others in the state
$(|a\rangle + |b\rangle)/\sqrt{2}$; (ii) Move the potential
corresponding to the internal level $|c\rangle$ back and forth so that
if the first atom is in that state it interacts with all the other atoms
for a time such that in each ``collision'' the phase shift difference
between the collision $a$--$c$ and $b$--$c$ is $\pi$; (iii) apply a
$\pi/2$ pulse (in transition $a$--$b$) to all the atoms except the first
one, which is transferred from $|c\rangle$ to $|b\rangle$. We emphasize that the optical lattice configuration allows the preparation of these states in a {\em single sweep} of the lattice. 
(e) Finally, it is
clear that one could use optical lattices in the context of quantum
information since the above procedure provides a fundamental two bit
gate (\ref{gate}) which, combined with single particle rotations allows
to perform any quantum computation between an arbitrary number of
two--level systems. In particular, the optical lattice setup
would be very well suited to implement fault tolerant quantum computations
due to the possibility of doing gate operations \cite{Gatepapers} in parallel (details will be given in  \cite{unpub}).

So far we have neglected some processes that may lead to decoherence, and
therefore limit the performance of our scheme. This includes spontaneous emission of atoms in the off-resonant optical lattice potentials, and inelastic collisions to wrong final atomic states. The spontaneous emission lifetime of a single atom in the lattice is from seconds to many minutes, depending on the laser detuning \cite{Hansch}. The problem of collisional loss is closely related to the loss mechanism in Bose condensates in magnetic and optical traps, a problem studied extensively in recent experiments. 
 By choosing proper internal hyperfine 
states one can maximize the lifetime due to inelastic collisions to be of the order of at least several seconds \cite{Ketterle}. 

Finally, by filling the lattice from
a Bose condensate, and using the ideas related to Mott transitions in optical lattices
\cite{Mott} it is possible to achieve uniform lattice occupation (``optical crystals'') or even specific atomic patterns,
as well as the  low temperatures necessary for performing the experiments
proposed in this letter. 


This research was supported 
by the Austrian Science Foundation, 
by the TMR network ERB-FMRX-CT96-0087, 
by the NSF under Grant No. PHY94-07194, and
by the Marsden fund, contract PVT-603.

{\em Note added:} After this work was completed we became aware of 
G. K. Brennen {\em et al.}, quant-ph/9806021, 
where  dipole--dipole interactions are proposed as a mechanism to 
implement a two bit quantum gate.



\begin{references}
\bibitem{foot3}
In $^{87}$Rb and $^{23}$Na one could use $|a\rangle\!\equiv \!|F=1,m_F\!=\!-1\rangle$,
$|b\rangle\!\equiv \!|F=2,m_F\!=\!2\rangle$ and $|c\rangle\!\equiv \!|F=\!1\!,m_F\!=\!1\rangle$ (see \protect \cite{unpub}).
\end{references}
\end{document}